\documentclass[doublespacing]{elsart}
\usepackage{stmaryrd}
\usepackage{amsmath}
\usepackage{amssymb}
\usepackage{amsfonts}
\usepackage{graphicx}
\usepackage{ctex}
\usepackage{rotating}
\usepackage{longtable}
\date{}
\begin{document}

\noindent \textbf{Corresponding author: }\\
 \noindent Dr. Hong-Jian Feng\\
\noindent
Department of Applied Physics,\\
School of Science,\\
Chang'an University,\\ Xi'an 710064, China\\
 Tel.:
+86-29-82337620\\
Email address:\\
fenghongjian@gmail.com\\
fenghongjian1978@gmail.com\\
 \clearpage

\begin{frontmatter}



\title{The reversal of magnetization driven by the Dzyaloshinskii-Moriya interaction(DMI) in perovskite Bi$_2$FeMnO$_6$}


\author{Hong-Jian  Feng }

\address{ Department of Applied Physics, School of Science, Chang'an University, Xi'an 710064, China}

\begin{abstract}
$Ab$ $initio$ calculations show that the coupling between
antiferrodistortive(AFD) distortions and magnetization  in
perovskite Bi$_2$FeMnO$_6$ is prohibited to make magnetization
rotate as on-site Coulomb interaction($U$) is larger than 2.7 eV,
where anomalies in antiferromagnetic(AFM) vectors and band gap
varying with on-site Coulomb interaction can be observed. This
coupling is attributed to the antisymmetric Dzyaloshinskii-Moriya
interaction(DMI) driven by the $e_g$-$e_g$ states AFM interaction
and charge redistribution with respect to different AFD distortions.

\end{abstract}

\begin{keyword}
Dzyaloshinskii-Moriya interaction(DMI);On-site Coulomb Interaction
;Bi$_2$FeMnO$_6$
\PACS 75.30.Et,75.30.GW,71.15.Mb
\end{keyword}
\end{frontmatter}


\section{ Introduction}
Multiferroics are materials in which ferroic properties, e.g.,
magnetism and polar order coexist. Magnetic and ferroelectric
ordering couple microscopically or macroscopically to form the
magnetic ferroelectrics. The coupling of the two ordering leads to
the so-called magnetoelectric effect in which the magnetization can
be tuned by the external electric field, and vice versa\cite{1,2,3}.
Magnetic ferroelectrics have potential applications in information
storage, actuators, sensors, and functional devices. Perovskite
BiFeO$_3$ exhibits both weak ferromagnetism and ferroelectric
characteristics, and has  been studied extensively in recent
days.\cite{4,5,6,7,8}. The G-type antiferromagnetic (AFM) order of
Fe magnetic moments exhibits a canting caused by the antisymmetric
Dzyaloshinskii-Moriya interaction(DMI) under rhombohedral  $R3c$
space group. However,  a spiral spin structure of AFM Fe sublattice
rotates  through the crystal with a long-wavelength period of
620{\AA}, and decrease the weak ferromagnetism further. In our
previous work, we suggested that the effect of  decreasing magnetism
caused by the spiral spin structure can be suppressed  by doping
magnetic transitional metal ions in perovskite B sites\cite{9}. We
suggest perovskite Bi$_2$FeMnO$_6$ is a good candidate to fulfill
this requirement in that Mn and Fe have different magnetic moments
and hence lower DMI to produce the rotation of AFM vectors.
Comparing with BiFeO$_3$, the new system should have an observed
magnetization and reversal of magnetization macroscopically. Except
the ferroelectric displacement of Bi site driven by  Bi-6s
stereochemically active lone pair induced by the mixing between the
$(ns)^2$ ground state and  a low-lying $(ns)^1(np)^1$ excited state,
there exists another structural distortion, the alternating sense of
rotation of the oxygen octahedra along [1 1 1] direction, which is
known as antiferrodistortive(AFD) distortion \cite{10,11}.  Also in
our former paper, it is shown that AFD distortion couples with the
weak ferromagnetism due to the DMI which is decreasing with
increasing on-site Coulomb iteraction($U$) \cite{12}. We aim to
extend our study to perovskite Bi$_2$FeMnO$_6$ to investigate the
coupling between AFD distortion and weak ferromagnetism under DMI
using $ab$ $initio$ calculations with considering the
spin-orbital(SO) coupling effect and the noncollinear spin
configuration, while Mn were doped periodically along [1 1 1]
direction in Fe site of BiFeO$_3$ .Does AFD distortion couple with
magnetism in Bi$_2$FeMnO$_6$ when Fe and Mn are arranged in G-type
AFM ordering? What is the origin of coupling between the AFD
distortion and the magnetism under DMI in Bi$_2$FeMnO$_6$? What is
the role of DMI in the coupling between AFD distortion and
magnetization? How does Coulomb iteraction($U$) take effect on the
DMI?  In this paper we have proposed the mechanism of DMI in
Bi$_2$FeMnO$_6$,using first-principles calculations based on density
functional theory(DFT).This work can shed light on discovering the
magnetoelectric coupling process in perovskite multiferroics.

The remainder of this paper is organized as follows: In section 2,
we presented the computational details of our calculations.  We
reported the calculated results and discussions in section 3. In
section 4, we drew  conclusion based on our calculation.

\section{ Computational details}

 Our calculations were performed within the local spin density
 approximation(LSDA) to DFT using the
ABINIT package\cite{13,14}.  The ion-electron interaction was
modeled by the projector augmented wave (PAW) potentials
\cite{15,16} with a uniform energy cutoff of 500 eV. Bi 5d, 6s, and
6p electrons, Fe 4s, 4p,and 3d electrons, and O 2s and 2p electrons
were considered as valence states. Two partial waves per $l$ quantum
number were used. The cutoff radii for the partial waves for Bi, Fe,
and O were 2.5, 2.3, 1.1 a.u., respectively. $6\times6\times6$
Monkhorst-Pack sampling of the Brillouin zone were used for all
calculations. We calculated the net magnetization per unit cell and
the electronic properties within the LSDA+U method where the strong
Coulomb repulsion between localized $d$ states has been considered
by adding a Hubbard-like term to the effective
potential\cite{17,18,19}. The effective Hubbard parameter, the
difference between  the Hubbard parameter $U$ and the exchange
interaction $J$ ($U-J$), was changing in the range between 0 and 5
eV for the Fe and Mn $d$ states. For 0 eV of $(U-J)$, $J$ was
varying as 0,0.5, 0.8,and 1 eV, respectively. $J$ remained 1 eV for
other effective Hubbard values. Taking into account the SO
interaction, we introduced the noncollinear spin configuration to
construct the G-type AFM magnetic order with the AFM axis being
along the $x$ axis in Cartesian coordinates  in our $ab$ $initio$
calculation.

\section{ Results and discussion}

The initial lattice parameters are introduced as same as BiFeO$_3$
in Ref.[4]. Then cell shape and the atomic positions are fully
relaxed, and the relaxed parameters are given in table 1. One can
see that the lattice constant is decreased with doping Mn in
BiFeO$_3$. The rhombohedral angle is also decreased comparing with
BiFeO$_3$. This shows that the lattice cell is compressed with
substitution of Mn in Fe site while lattice shape is also changed
with decreasing rhombohedral angle.

For the AFD distortion, a rotational vector $\mathbf{R}$ has been
introduced to describe the direction of the rotation of the oxygen
octahedra\cite{12}. $\mathbf{R}_{out}$ corresponds to the state in
which the rotational vectors of two neighboring oxygen octahedra are
deviating away, while $\mathbf{R}_{in}$ is pointing inward. The
rotational angle is 10\textordmasculine   in the Cartesian
coordinates\cite{12}. $U$ and $J$ are the Coulomb interaction and
superexchange interaction parameter which were implemented in LSDA+U
calculation as in ref.[12]. $U$ is the amount of energy required to
transfer one electron from one site to its nearest neighbor, while
$J$ indicates the strength of the magnetic superexchange
interaction.

In table 2, we present the net magnetization per unit cell with
respect to $\mathbf{R}_{in}$ and $\mathbf{R}_{out}$ in Cartesian
coordinates for different $U$ and $J$. For the sake of clarity, we
only set varying value of  $J$  for $U$=0 eV. Again, one can  see
that $J$ value have no effect on the resulting magnetization when
$U$ remains constant. The initial magnetic moment of Fe are arranged
in ferrimagnetic order along $x$ direction which is known as G-type
AFM order. The resultant magnetic moment in $x$ direction  is due to
the ferrimagnetic arrangement of Fe and Mn magnetization.  The
neighboring moment will interact under the DMI and deviate away from
the original $x$ direction to produce a resultant magnetic moment in
$y$ direction. When the rotational vector $\mathbf{R}$ is reversed,
the resultant magnetic moment will be reversed too until the $U$ is
smaller than 2.7 eV. This indicates the AFD and magnetic ordering is
coupled and DMI is still active in this case. As $U$ is greater than
2.7 eV, the coupling between these two ordering is prohibited. That
is to say, the DMI is precluded in this case. It is worth mentioning
that the component of magnetic moment in $y$ direction for U =2.7 eV
becomes zero. This indicates the critical value of $U$ for
depressing DMI is 2.7 eV in which the antisymmetric Fe and Mn
magneitc moments stop interacting with each other. This critical
value is smaller than that for BiFeO$_3$ where the DMI is turned off
as $U$ is approaching 2.9 eV\cite{12}. This indicates that  Fe-Mn
antisymmetric interaction where interacting spin moment are
$(5/2)S-(4/2)S$ is weaker than Fe-Fe antisymmetric interaction where
interacting spin moment are $(5/2)S-(5/2)S$. We believe the weaker
DMI  and alternating spin moments in the new system would destroy
the spiral spin structure through the crystal in comparison with the
case in BiFeO$_3$. Hereby the magnetization and reversal of
magnetization can be improved in the new system. In order to make
the reversal of magnetization clear, the resultant AFM vectors are
illustrated in Fig.1 with respect to $U$. AFM vectors have three
components corresponding to $x$, $y$, and $z$ direction,
respectively. It is apparent that there is an anomaly as $U$ is
approaching 2.7 eV which corresponds to the critical value of
eliminating DMI. The magnetization is inverted as $U$ is under the
critical value which indicates the DMI is still active in this $U$
range. Meanwhile the DMI is inversely proportional to the $U$ value.
The deviation of original magnetic moment away from $x$ axis only
occurs in $xoy$ plane. That is why the reversal of magnetic moment
can not be observed in $z$ direction.

The relationship between DMI and Coulomb on-site interaction can be
understood theoretically as follows.  The Hamiltonian of the new
perovskite system reads,
\begin{equation}
H_{Bi_2FeMnO_6}=-2\sum_{<Fei,Mnj>}\textbf{J}_{Fei,Mnj}\mathbf{S}_{Fei}\cdot\mathbf{S}_{Mnj}+\sum_{<Fei,Mnj>}\textbf{D}_{Fei,Mnj}\cdot\mathbf{S}_{Fei}\times\mathbf{S}_{Mnj}.
\end{equation}
The first term is attributed to  the symmetric superexchange, and
the second term is the antisymmetric DMI contribution which only
occurs when the inversion symmetry of the cations is broken. The
constant \textbf{D} is related to the AFD displacement and  the
rotational vector. It functions like a 'DMI constant' which
indicates the strength of the DMI. It reads by the second order
perturbation in the case of one electron per ion
\begin{equation}
\textbf{D}_{Fe,Mn}^{(2)}=(4i/U)[b_{nn'}(Fe-Mn)C_{n'n}(Mn-Fe)-C_{nn'}(Fe-Mn)b_{n'n}(Mn-Fe)],
\end{equation}
From this relation, it can be seen that the DMI constant is
inversely proportional to the on-site Coulomb interaction and has
nothing to do with the symmetric superexchange parameter.

The new system should possess a good insulating property in terms of
the band gap calculated, and hence a reversal states of
ferroelectricity. The band gap for metal oxides can be opened with
adopting an appropriate $U$ value. The corresponding value for
BiFeO$_3$ is about 4.3 eV. We take 5 eV $U$ value for Fe and Mn in
Bi$_2$FeMnO$_6$ to analyze the total density of states(DOS)  in that
the band gap can be increased slightly with increasing $U$ value.
The band gap corresponding to two rotational vectors are
demonstrated in Fig.2. Generally, the band gap corresponding to the
two rotational vectors are increasing linearly with respect to $U$
value, while there exist two anomalies to the two AFD states as $U$
approaching  2.7 eV where the band gap are disappeared. These
anomalies in the curve are consistent with the calculating results
for AFM vectors  shown in table 2 and Fig. 1, respectively. The
anomalies show the disappearance of DMI occurs at the critical $U$
value and is caused by the hopping of electrons through Fermi
energy. Moreover, the band gap to $\mathbf{R}_{out}$ is more
narrower than to $\mathbf{R}_{in}$ which is probably caused by the
different ligand field related with structure. In comparison with
the total DOS without applying the on-site Coulomb interaction, the
total DOS with $U$ value of  5 eV is shown in Fig. 3. The finite DOS
in the vicinity of Fermi level is pushed away to form the band gap
as $U$ is equal to 5 eV. A 1.26 eV and 0.75 eV band gap is obtained
for $\mathbf{R}_{out}$ and $\mathbf{R}_{in}$ rotational states,
respectively. This indicates that these two AFD states both possess
a semiconducting property. A ferroelectric reversal would be
expected to these two reversal AFD states. However, Bi$_2$FeMnO$_6$
remains metal property without applying $U$. Again this  shows that
the on-site Coulomb interaction is significant in opening the band
gap for perovskite metal oxides.

In order to shed light on the role of DMI between neighboring Fe and
Mn ions, we report the orbital resolved density of states(ODOS) of
the magnetic ions as $U$ is equal to 2.6 eV which is exactly under
the critical value for precluding the antisymmetric DMI. The ODOS of
Fe and Mn corresponding to $\mathbf{R}_{out}$ and $\mathbf{R}_{in}$
states are shown in Fig. 4.  It is apparent that there are large
amount of doubly degenerate $e_g$ states for Fe and Mn in the
occupied bands around the vicinity of the Fermi level. The charge
hybridization between Fe-3d and Mn-3d electrons is realized mainly
by the $e_g$-$e_g$ states interaction. It suggests that the
$e_g$-$e_g$ AFM interaction play an important role in the
antisymmetric DMI between Fe and Mn ions. $e_g$ state is composed by
$d_{z^2-r^2}$ and $d_{x^2-y^2}$ orbitals. It is worth mentioning
that the finite ODOS in the vicinity of Fermi level is mainly
attributed to the $d_{z^2-r^2}$ orbital. We suggest that $e_g$-$e_g$
AFM interaction is exactly mediated by neighboring $d_{z^2-r^2}$
orbitals of Fe and Mn ions. The charge transformation between Fe and
Mn ions, and hence the inversion of magnetization can be seen from
Fig. 5. The hopping of electrons between neighboring $e_g$-$e_g$
pairs occurs due to the DMI as the inversion symmetry of cations is
broken. Moreover the hopping of electrons couples with the rotation
of the neighboring oxygen octahedra to produce the resultant
magnetization. The charge distribution in neighboring $e_g$-$e_g$
states is polarized to a certain rotational vector, and the charge
distribution is inversely polarized as the rotational vector is
changing direction, leading to the reversal of resultant
magnetization eventually. The charge distribution is supposed to
distribute uniformly without applying  the DMI. However, the
homogenous distribution is destroyed by the DMI between neighboring
magnetic ions to form a polarized status. We suggest this
constitutes the microscopic mechanism of DMI in perovskite
multiferroics. On the other hand, the AFM interaction in triply
degenerate $t_{2g}$-$t_{2g}$ composed by $d_{xy},d_{yz},$ and
$d_{xz}$ orbitals is relatively weak comparing with $e_g$-$e_g$ AFM
interaction. Again this shows that the latter is deeply involved in
the antisymmetric DMI.

\section{ Conclusion}
The critical value of $U$ for depressing DMI in Bi$_2$FeMnO$_6$ is
found to be 2.7 eV which is smaller than that in BiFeO$_3$. This
indicates that the DMI in the new system functions weaker in
comparison with that in BiFeO$_3$ .  ODOS of Fe and Mn corresponding
to different AFD displacements show that the microscopic mechanism
of the DMI is originating from the hopping of electrons between
neighboring $e_g$-$e_g$ states, and the $e_g$-$e_g$ AFM interaction
couples with the rotation of the neighboring oxygen octahedra.



\clearpage

\begin{table}[!h]

\caption{ Calculated lattice constant a, rhombohedral angle
$\alpha$, volume V, and Wyckoff parameters for Bi$_2$FeMnO$_6$. The
Wyckoff positions 1a(x,x,x) for the cations and 3b(x,y,z) for the
anions.}

\begin{center}

\tabcolsep=8pt
\begin{tabular}{@{}ccccccc}
\hline\hline
  &&Bi$_2$FeMnO$_6$   &BiFeO$_3$&\\
\hline
a({\AA})&&  5.030& 5.459\\
$\alpha$(\textordmasculine )&&59.02&60.36\\
V({\AA$^3$})&& 105.57  & 115.98 \\
Bi& x& -0.012/0.489 & 0.000   \\
Fe/Ti& x& 0.226(Fe)/0.727(Mn) & 0.231  \\
O&x& 0.541 & 0.542  \\
&y& 0.950 & 0.943  \\
&z&  0.406  & 0.398 \\
\hline\hline
       \end{tabular}

       \end{center}
       \end{table}

\begin{table}[!h]

 \caption{ Magnetization per unit cell with respect to different value of $U$ and$J$.}

\begin{center}

\begin{tabular}{@{}ccccccccccc}
\hline\hline
$U(eV)$&\multicolumn{2}{c}{0}&\multicolumn{2}{c}{0.5}&\multicolumn{2}{c}{0.8}&
\multicolumn{2}{c}{1}&\multicolumn{2}{c}{2}\\
$J(eV)$&\multicolumn{2}{c}{0}&\multicolumn{2}{c}{0.5}&\multicolumn{2}{c}{0.8}&
\multicolumn{2}{c}{1} &\multicolumn{2}{c}{1}\\
&$\mathbf{R_{in}}$&$\mathbf{R_{out}}$&$\mathbf{R_{in}}$&$\mathbf{R_{out}}$&$\mathbf{R_{in}}$&
$\mathbf{R_{out}}$&$\mathbf{R_{in}}$&$\mathbf{R_{out}}$&$\mathbf{R_{in}}$&$\mathbf{R_{out}}$\\
\hline
$M_x(\mu_B)$&-2.9723&1.0683&-2.9723&1.0683&-2.9723&1.0683&-2.9723&1.0683&-2.9775&1.0041\\
$M_y(\mu_B)$&-0.0079&0.0887&-0.0079&0.0887&-0.0079&0.0887&-0.0079&0.0887&-0.0100 &0.0367\\
$M_z(\mu_B)$&0.0004&0.0071&0.0004&0.0071&0.0004&0.0071&0.0004&0.0071&0.0021&0.0025\\
\hline\hline
$U(eV)$&\multicolumn{2}{c}{3}&\multicolumn{2}{c}{3.5}&\multicolumn{2}{c}{3.6}&
\multicolumn{2}{c}{3.7}&\multicolumn{2}{c}{3.8}\\
$J(eV)$&\multicolumn{2}{c}{1}&\multicolumn{2}{c}{1}&\multicolumn{2}{c}{1}&\multicolumn{2}{c}{1}&
\multicolumn{2}{c}{1}\\
&$\mathbf{R_{in}}$&$\mathbf{R_{out}}$&$\mathbf{R_{in}}$&$\mathbf{R_{out}}$&$\mathbf{R_{in}}$&
$\mathbf{R_{out}}$&$\mathbf{R_{in}}$&$\mathbf{R_{out}}$&$\mathbf{R_{in}}$&$\mathbf{R_{out}}$\\
\hline
$M_x(\mu_B)$&-2.9813&1.0159&-2.9843&1.0126&-2.9846&1.0108&-2.9851&1.0087&-2.9856&1.0118\\
$M_y(\mu_B)$&-0.0050&0.0208&-0.0013&0.0194&-0.0004&0.0205&0.0000&0.0211&0.0002&0.0238\\
$M_z(\mu_B)$&0.0021&-0.0008&0.0013&-0.0004&0.0011&0.0001&0.0009&0.0009&0.0003&0.0017\\
\hline\hline $U(eV)$&\multicolumn{2}{c}{3.9}&\multicolumn{2}{c}{4}&
\multicolumn{2}{c}{5}&\multicolumn{2}{c}{6}\\
$J(eV)$&\multicolumn{2}{c}{1}&\multicolumn{2}{c}{1}&
\multicolumn{2}{c}{1}&\multicolumn{2}{c}{1}\\
&$\mathbf{R_{in}}$&$\mathbf{R_{out}}$&$\mathbf{R_{in}}$&$\mathbf{R_{out}}$&
$\mathbf{R_{in}}$&$\mathbf{R_{out}}$&$\mathbf{R_{in}}$&$\mathbf{R_{out}}$&$\mathbf{R_{in}}$&$
\mathbf{R_{out}}$\\
\hline
$M_x(\mu_B)$&-2.9862&1.0115&-2.9866&1.0115&-2.9919 &1.0093  &-2.9921 &1.0074\\
$M_y(\mu_B)$&0.0007&0.0231&0.0011&0.0233&0.0090&0.0220& 0.123&0.0144 \\
$M_z(\mu_B)$&0.0003&0.0025&0.0000&0.0025&0.0000&0.0018&-0.0003& 0.0013  \\
 \hline\hline
\end{tabular}

\end{center}

\end{table}

\clearpage

\raggedright \textbf{Figure captions:}

Fig.1 Relationship between AFM vectors and $U$ .

 Fig.2  Band gap as a function of $U$.

 Fig.3 Total DOS for \textbf{R}$_{in}$ and \textbf{R}$_{out}$ with
$U$=5 eV and without applying $U$.

 Fig. 4 ODOS of Fe and Mn for \textbf{R}$_{in}$ and
\textbf{R}$_{out}$ with $U$=2.6 eV and without applying $U$.

 Fig. 5 Schematic diagram for rotation of magnetization and DMI between the neighboring Fe and Mn
ions  in Bi$_2$FeMnO$_6$. The arrow denote the magnetic moment.

\clearpage
\begin{figure}
\centering
\includegraphics{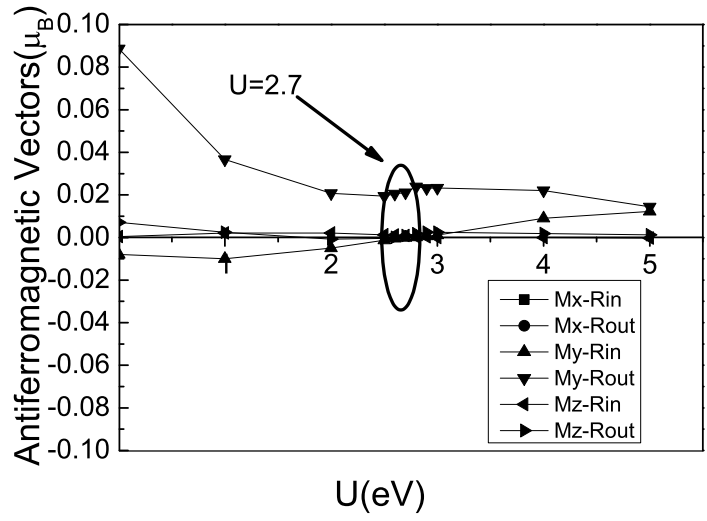}
\caption{Relationship between AFM vectors and $U$ .}
\end{figure}

\begin{figure}
\centering
\includegraphics{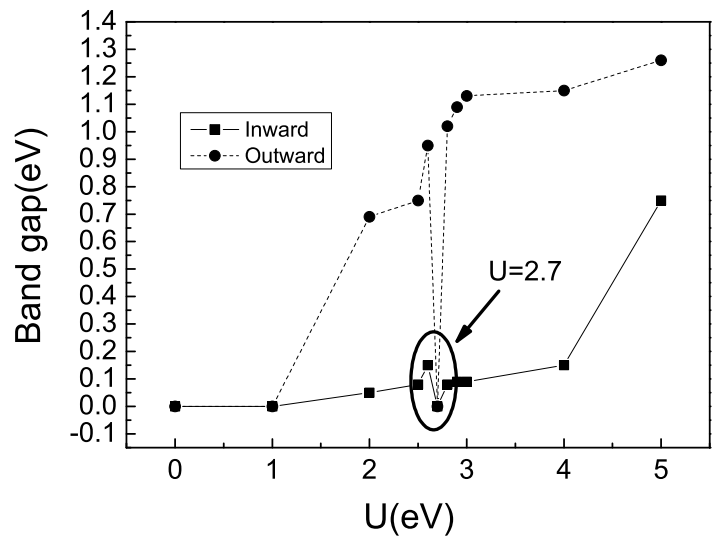}
\caption{ Band gap as a function of $U$.}
\end{figure}

\begin{figure}
\centering
\includegraphics{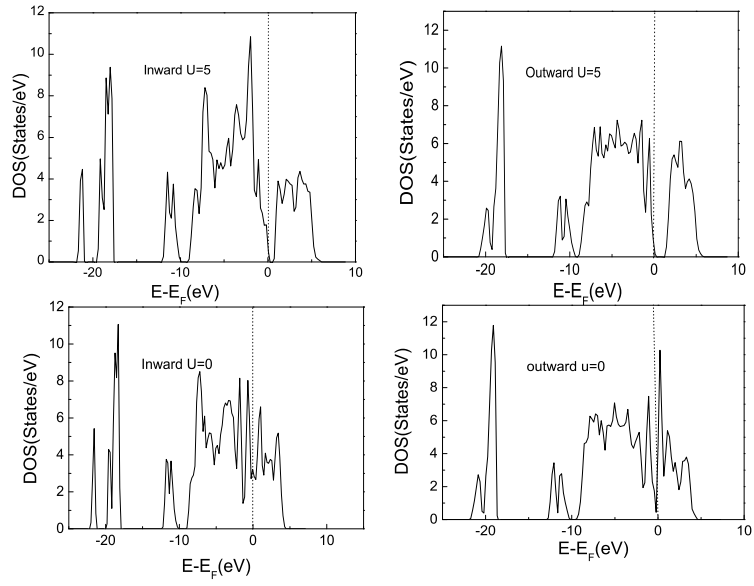}
\caption{Total DOS for \textbf{R}$_{in}$ and \textbf{R}$_{out}$ with
$U$=5 eV and without applying $U$. }
\end{figure}

\begin{figure}
\centering
\includegraphics{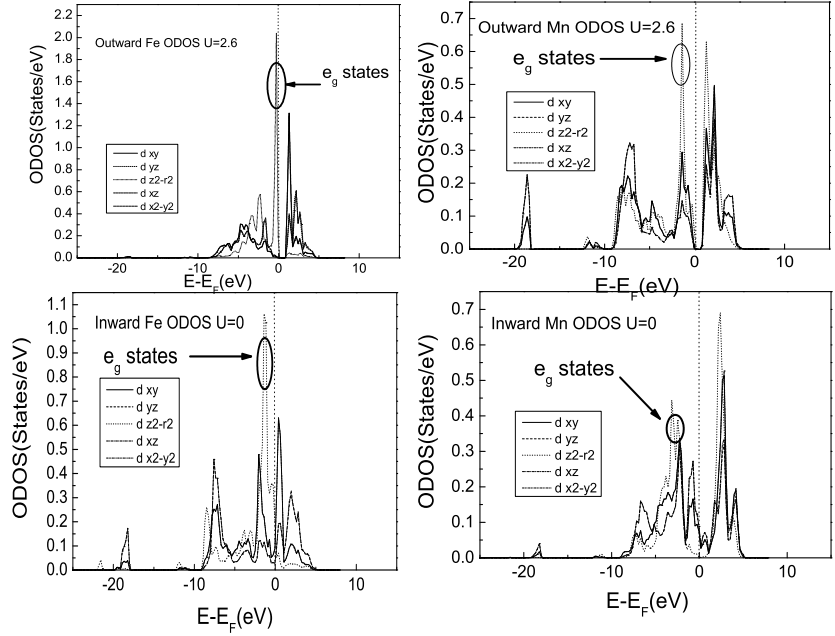}
\caption{ODOS of Fe and Mn for \textbf{R}$_{in}$ and
\textbf{R}$_{out}$ with $U$=2.6 eV and without applying $U$.}
\end{figure}

\begin{figure}
\centering
\includegraphics{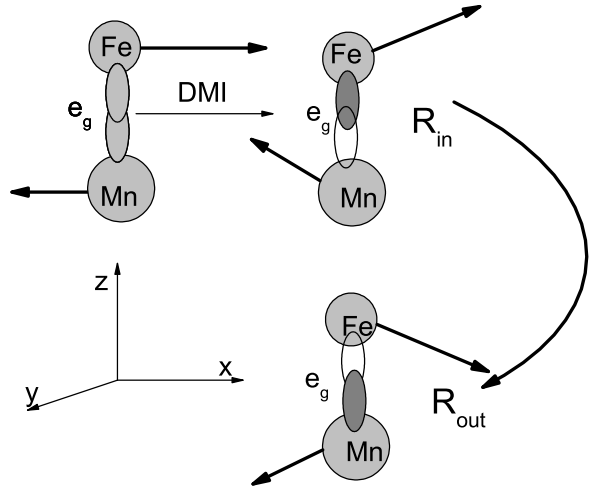}
\caption{Schematic diagram for  rotation of magnetization and DMI
between the neighboring Fe and Mn ions  in Bi$_2$FeMnO$_6$. The
arrow denote the magnetic moment.}
\end{figure}

\end{document}